# Suitability of NVIDIA GPUs for SKA1-LOW

**Alessio Magro[1,E], Kristian Zarb Adami[1,2], Steve Ord[3]**

[1] Institute of Space Science and Astronomy (ISSA), University of Malta, Malta
[2] University of Oxford, UK
[3] ICRAR, Curtin University, Australia
[E] Author email: alessio.magro@um.edu.mt

# 1. Introduction

In this memo we investigate the applicability of NVIDIA Graphics Processing Units (GPUs) for SKA$_1$-low station and Central Signal Processing (CSP)-level processing. Station-level processing primarily involves generating a single station beam which will then be correlated with other beams in CSP. Fine channelization can be performed either at the station or CSP-level, while coarse channelization is assumed to be performed on FPGA-based Tile Processors, together with ADC conversion, equalization and other processes. Rough estimates for number of GPUs required and power requirements will also be provided.

# 2. GPU and Algorithmic Performance

Two NVIDIA GPUs will be considered in this memo, the Tesla K40 and GTX 750 Ti, the former being a high-performance Kepler card whilst the latter is the first Maxwell-series power efficient card. The technical specifications of both cards are listed in Table 1.

|  | **K40** | **750 Ti** |
|---|---|---|
| SP Peak performance (GFLOPS) | 4.29 | 1.3 |
| GDDR5 Memory (GiB) | 12 | 2 |
| PCI-E version | 3 | 3 |
| Internal memory bandwidth (GB/s) | 288 | 86.4 |
| Max TDP (W) | 235 | 60 |
| Dual copy engine | Yes | No |

Table 1. NVIDIA Tesla K40 Specifications

Table 2 lists the processing performance for beamforming, correlation and channelisation, together with their implementation efficiency, as defined in [1,2]. Although the cross-correlation results were measured on an older Fermi-based device, we keep the measured peak performance of 78%. Beamforming performance was measured on an NVIDIA Tesla K20. Channelisation performance is based on an implementation prototype which was benchmarked on a K20, as well as benchmarking of the cuFFT library. Specialised implementations can probably achieve a higher efficiency. PCI-e transfer efficiency is assumed to be 85%.

|  |  |  | **Efficiency** |
|---|---|---|---|
| Beamformer input | $S_{in}$ | $2B\eta N_a N_p n_s$ | 85% |
| Channelisation | $C_{ops}$ | $BN_b N_a N_p \times (\log_2(N_c) + 8N_{taps})$ | 50% FIR, 30% FFT |
| Cross-Correlation | $X_{ops}$ | $8 \times \frac{1}{2} BN_b N_p N_s (N_p N_s + 1)$ | 78% |
| Beamforming | $B_{ops}$ | $8 \times BN_b N_a N_p$ | 60% |
| Beamformer output | $B_{out}$ | $2B\eta N_b N_p n_b$ | 80% |
| Correlator output | $X_{out}$ | $2B\eta \times \frac{1}{2} N_p N_a (N_p N_a + 1) \times 1/\tau_{dump} \times n_c$ | 85% |

Table 2. Process operation count

- $2B$ is the Nyquist sampled bandwidth
- $N_b$ is the number of beams
- $N_a$ is the number of antennas
- $N_s$ is the number of stations
- $N_p$ is the number of polarisations
- $N_{taps}$ is the number of taps for an assumed polyphase filterbank channeliser
- $\eta$ is the coding factor (assumed to be 1.25)
- $\tau_{dump}$ is the correlator dump time
- 85% PCIe efficiency is assumed (where maximum PCIe 3 bandwidth is 15.75 GB/s)
- $n_s$ is the input sample bitwidth to the beamformer
- $n_b$ is the beam output sample bitwidth
- $n_c$ is the correlation output sample bitwidth

## 3. SKA$_1$-Low Station Processing

The system specification specified in the SKA$_1$ system baseline design [3] will be used to compute the GPU requirements for station-level processing:

- Each of the 911 stations is composed of 289 dual-pol antennas
- Each antenna signal is digitised at 500 MSA/s, resulting in 250 MHz of processable bandwidth
- An FPGA-based, 1024-point PFB performs coarse channelisation on every antenna/polarisation
- Data is transferred over 10 Gb/s links, requiring a minimum of 289 links (one per antenna)
- 40 Gb/s switches will be used to transfer data from the antennas to the processing servers
- Assuming appropriate data packetisation, each 40 Gb/s link will carry all dual-pol antenna signals for 8 frequency channels
- Each GPU will process a subset of the frequency channels, for all antennas and both polarisations
- One beam needs to be generated per station
- Optionally, finer channelisation is performed at the station level, although this scheme might not be ideal for certain KSPs or observations
- Data is sent back to the switch, which will be transorted to the CSP

Performing channelisation after beamforming greatly reduces the processing requirements since $N_b \ll N_a$, and can thus be performed either at station or CSP level. When beamforming on GPUs it is advantageous to perform this at station level:

- Beamforming systems where a small number of beams need to be generated are PCIe bandwidth limited, resulting in idle GPU cores
- For a CSP FX correlator, data will need to be corner turned between the F and X stages. This corner turn can be performed automatically by the network switches connecting the station and CSP processing nodes if proper packetisation is used.

Table 3 lists the data transfer and processing requirements per coarse frequency channel. This results in a total input bandwidth requirement of ~336.44 GB/s and processing requirements of 1.15 TFLOPs. Figure 1 shows the resource utilisation of a single GPU when processing a subset of the frequency channels, based on device specifications for the K40, taking into account the efficiency factors listed in table 2. $N_{taps}$ is assumed to be 8. The system is clearly PCIe bandwidth limited, with each GPU capable of processing 40 frequency channels while using less than 4% of the total available processing power, assuming that PCIe I/O and kernel execution can be performed simultaneously (using two CUDA streams). A total of 26 GPUs per station would thus be required (23,686 GPUs for all stations).

| | |
|---|---|
| Input data rate | 336.44 MB/s |
| Beamforming | 1.13 GFLOPS |
| Channelisation | 0.08 GFLOPS |
| Output data rate | 1.16 MB/s |

Table 3. Data transfer and processing requirements per coarse channel for each SKA1-low station

Figure 2 plots the number of GPUs required for a varying number of input antennas and output beams for SKA1-low station parameters. The entire parameter space is bandwidth limited.

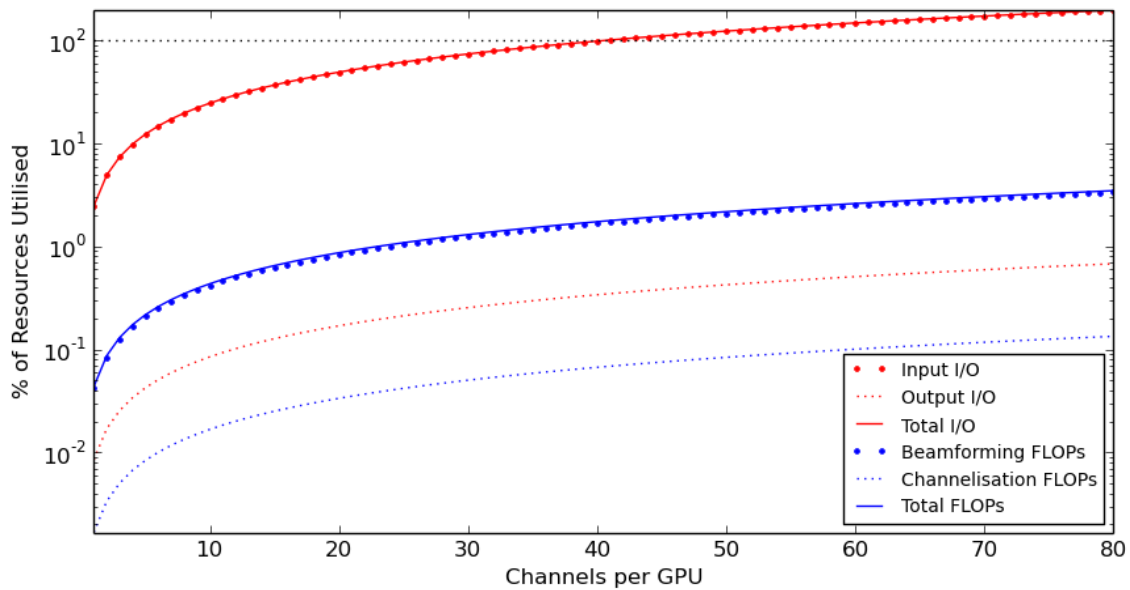

Figure 1. Resource utilisation for a GPU-based station-level beamformer and channeliser, based on specifications for the NVIDIA K40. The process is PCIe bandwidth limited, and ~40 channels can be processed in real-time on each GPU.

The K40 is not a suitable candidate for station-level processing. Desired features for a GPU-based design would include a GPU which:

- Supports PCIe 3 for maximum PCI bandwidth
- GDDR5 RAM for fast internal bandwidth
- Has a peak theoretical performance of at least 100 GFLOPS (multiplied by the discrepancy factors of internal memory bandwidths from the tested devices)
- As much RAM as possible, to reduce the number of iterations required per second
- Does not consume a lot of power
- Supports concurrenct data transfer and kernel execution

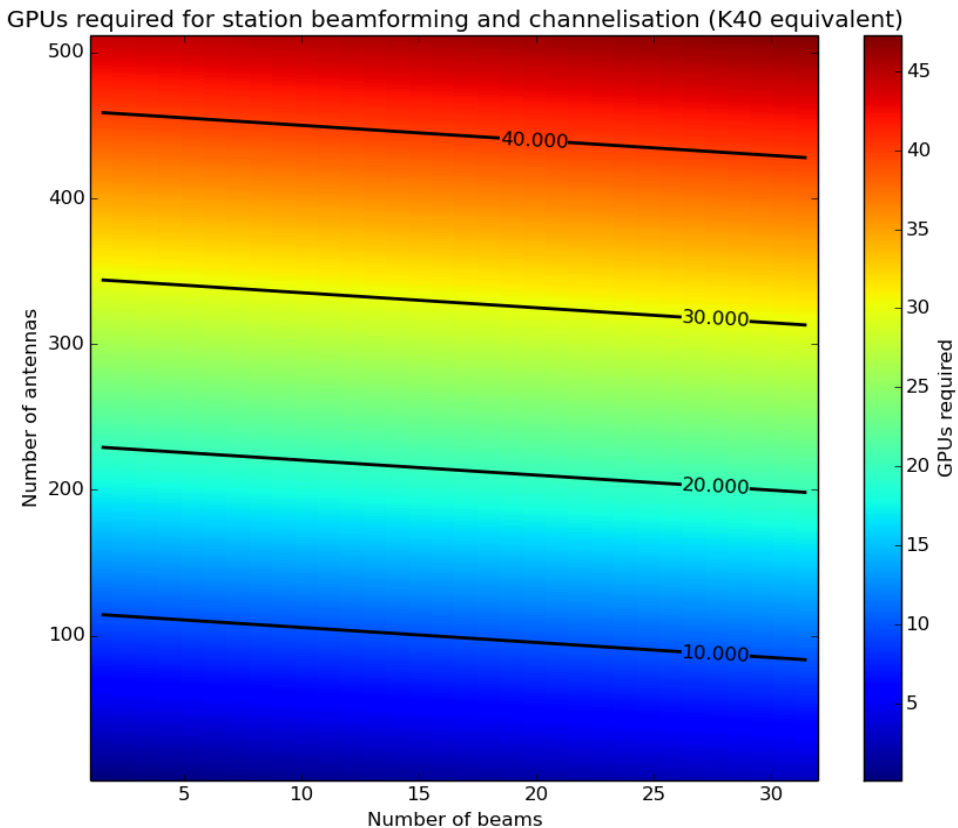

Figure 2. Station GPU requirements for varying number of input antennas and output beams for SKA1-low station parameters for beamforming and coarse channelisation, with factors of 10 delineated by solid black contour lines. The entire parameter space is PCIe bandwidth limited.

The new generation Maxwell cards, specifically the GTX 750 Ti, support the above minimum requirements of relatively low processing power and high power efficiency (even though they lack high speed internal RAM and have limited memory available), thus resulting in much lower deployment and running costs, with beamforming consuming about 20 W per card. Beamforming and channelization on these cards is still PCIe bandwidth limited, with less than 10% of available processing power being utilized, resulting in the same number of required GPUs.

Assuming that PCIe 3 will be the best available GPU interface, each GPU will transfer ~110 Gbps to/from the host servers. The currently available data transfer technologies include 40 Gb/s and 100 Gb/s ethernet, as well as 56 Gb/s Infiniband (higher speed Infiniband technologies will be availble by 2016). Assuming 40 Gb/s links, then dual-GPU servers would require 6 Ethernet links, or three dual-interface adapters, using a total of 5 PCIe 3 x16 slots. Each 40 Gb/s link transfer the data of 10 coarse frequency channel for all antennas and polarisations, precisely matching the GPU's transfer rate. The overall system architecture per station would consist of:

- 13 dual-GPU servers having a total of six 40 Gb/s Ethernet links. Each server should ideally have at least the CPU computational power equivalent to two quad-core CPUs for network stream processing (one core per 40 Gb/s link) and two cores for GPU management. A minimum of 64 GB high speed RAM is also required.

- 40 Gb/s network switch with 289 10 Gb/s input ports and 78 40 Gb/s output ports (with a small number of additional ports for data transport to the CSP and for managment and control links, if required). Since the input and output data flow are fully partitioned, this block can be composed of multiple switches requiring minimal cross-communication.

Additional notes:
- The above data rates exclude management and control
- The total GPU power consumption for the above setup is ~470 kW.
- Redundancies are not incorporated into these estimates
- To avoid unnecessary requantisation and increased memory footprint we have assumed the availability of native 8-bit support in the GPU-based channelisation. We anticipate support of this feature in the 2014 timeframe in NVIDIA's cuFFT library with CUDA 6.5.
- Memory limitations are not included in the above estimates
- The processing requirements for station processing could probably be handled by CPUs rather than GPUs. Currently a high-end CPU has a PTP of ~300 GFLOPS, so computation wise only about 5 would be required (assuming 100% utilisation rate and excluding data interpratation, data movement in memory, re-packetisation and so on). However, by 2016 CPUs will be capable of performing all this in real time on the same number of servers, significantly reducing deployment and running costs.
- The overall system is clearly bandwidth limited, suggesting that a PC/GPU-based architecture is not a viable option. By comparison, FPGA-based digital boards can generally process much higher bandwidths at lower running costs.
- NVIDIA's newly announced NVLink technology suggests that this bottleneck can be alleviated on setups which support this technology, however a clear roadmap for consumer harwdare is no yet available, and deployable systems might not be available by SKA1's technology freeze time frame.

## 4. SKA$_1$-Low Correlator

The following assumptions will be used to compute the GPU requirements for CSP correlation processing:

- SKA1-low will be composed of 911 stations
- Signals are pre-channelised into 262,144 fine frequency channels
- A single complex beam is output per station, having a bandwidth of 250 MHz

Following the above assumptions Table 3 lists the data transfer and processing requirements per fine frequency channel. This results in a total bandwidth requirement of ~1 TB/s and processing requirements of 3.3 PFLOPs. Figure 3 shows the resource utilisation of a single GPU when processing a subset of the frequency channels, based on device specifications for the K40, taking into account the efficiency factors listed in table 2. Cross-correlation is compute bound, with each GPU capable of processing 276 fine frequency channels, with less than 8% of PCIe resources being utilised. A total of 950 GPUs would thus be required to cross-correlate all station beams. Figure 4 shows the number of GPUs required for varying bandwidth and number of stations for SKA1-low CSP parameters. Most of the parameters space is compute bound, except for the area beneath the white contour line. Clearly, the major requirement for CSP GPUs is to have a high peak theoretical performance.

| Input data rate | 4.14 MB/s |
|---|---|
| Cross-Correlation | 12.68 GFLOPS |
| Output data rate | 0.00 MB/s |

Table 4. Data transfer and processing requirements per fine channel for the SKA1-CSP correlator

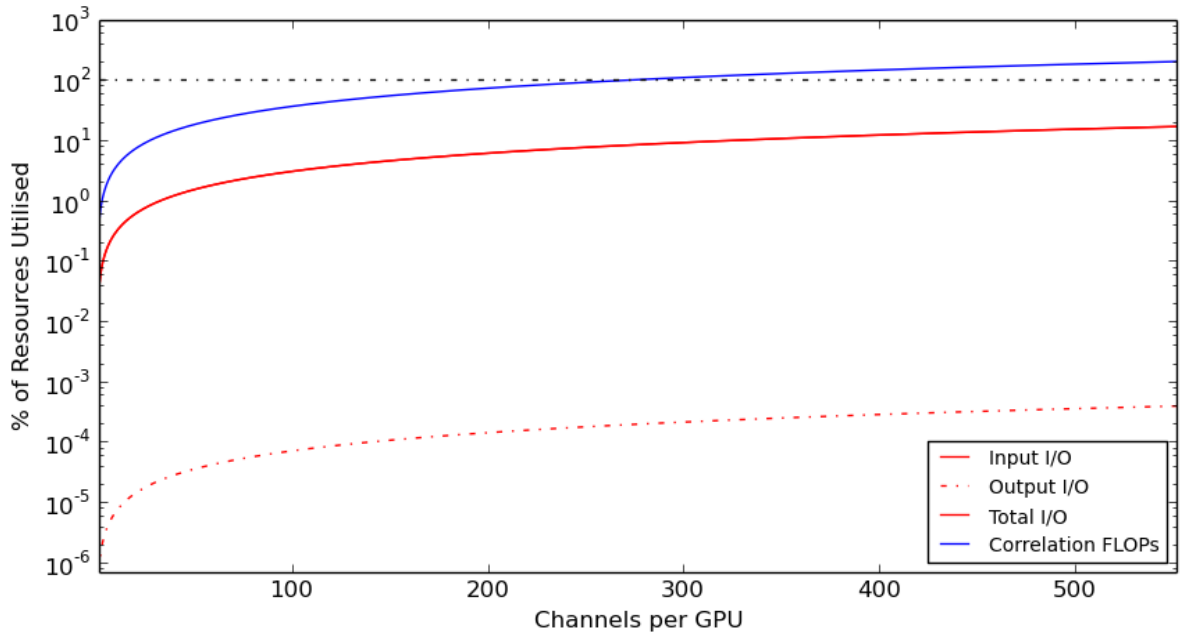

Figure 3. Resource utilisation for a GPU-based CSP corss-correlator, based on specifications for the NVIDIA K40. The process is compute limited, where ~276 channels can be processed in real-time on each GPU.

It follows from Figure 3 that a single K40 can perform the X part of an FX correlator for all fine frequency channels belonging to a single coarse subband for all stations, with enough computational leftover to integrate the F part of the correlator, resulting in the following compute utilization in each GPU:
- 15% for F part (assuming 32-bit FFTs)
- 72% for X part
- 13% left for corner turn (this is entirely bandwidth limited, and it's running time will be greatly reduced if Pascal GPUs manage to achieve the predicted 1 TB/s internal memory bandwidth)

This would reduce the computational requirements for station-level beamforming while retaining the same number of processing nodes in the CSP. For dual-GPU systems two 10 Gb/s links are required to feed the GPUs, whilst minimal CPU computational power is required. The CSP overall system architecture would therefore consist of:

- 512 dual-GPU servers with 2 10 Gb/s Ethernet links (which can be integrated into the motherboard). Each server requires minimal CPU power and should at least have double the total amount of GPU RAM (if all of it is being utilized). A single 6-core CPU and ~32 GB of (ideally) high speed RAM should suffice
- A 10 GbE network switch, or set of switches, having approximately 2048 ports.

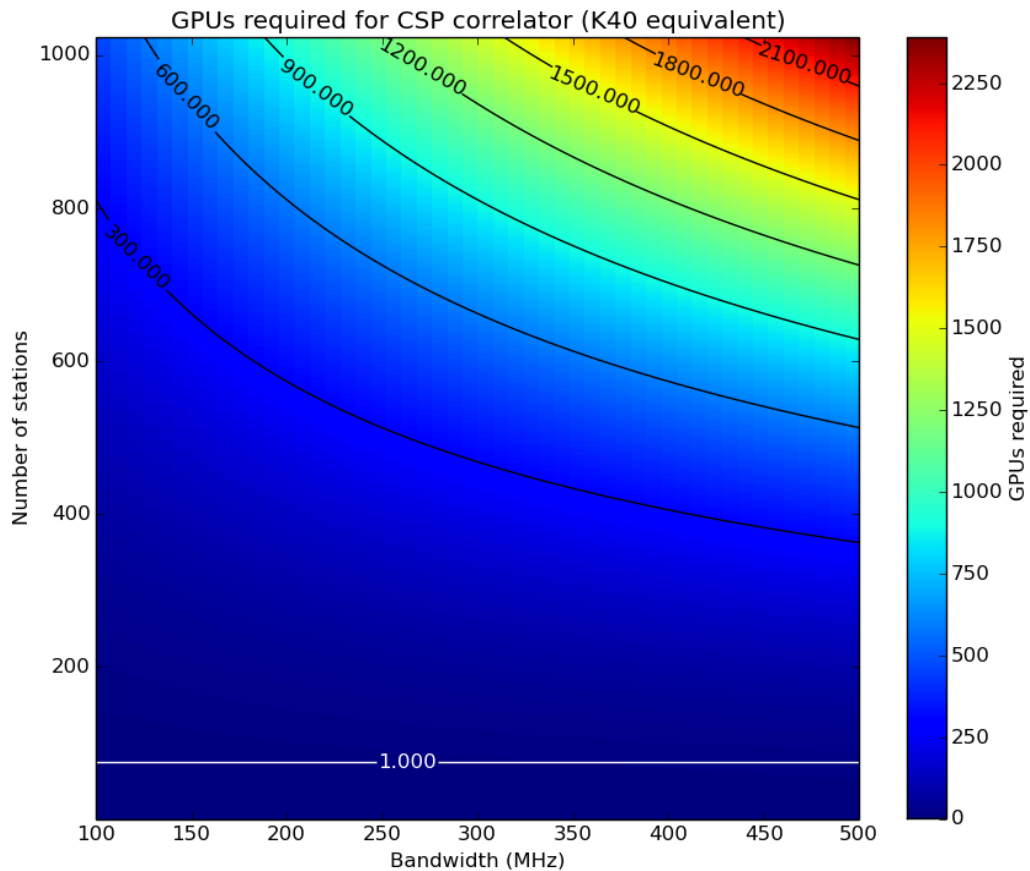

Figure 4. Number of GPUs required for varying bandwidth and number of stations for SKA1-low CSP parameters. Solid black contour lines partition the parameter space, which is bandwidth limited below the white contour line, while the rest is compute limited.

Each host, containing 2 K40s, can be package as a 1U blade. Assuming a rack size of 42U, and including rack space and 32 processing nodes per blade (leaving enough space to host switches, power distribution units and any additional hardware which might be required), a total of 16 racks are needed. Assuming a maximum GPU TDP of 235 W (which is the listed specification for the K40) and the host consuming ~200 W, the power envelope for this system would be ~22 kW, with a total power envelope of ~335 kW, excluding switches, cooling and additional hardware in the deployed system. These power estimates assume that correlation will pull all available GPU power, however it is predicted that by 2016, when the Pascal architecture is predicted to be released, the flop per watt ratio will be approximately 35 (ignoring further improvement such as careful GPU tuning, assembly-level tuning and using lower-bit multiplication). With this figure, the predicted power consumption on Pascal architectures for correlating all of SKA$_1$-low, assuming two GPUs per host processing the same number of fine frequency channels and similair host power consumption, is ~200 kW.

Additional Notes:
- It is assumed that by 2016 GPU processing power will at least double, so the estimated number of GPUs will diminish drastically, at least by a factor of 2, with a proportional linear reduction in power requirements. The predicted power consumption for this case is ~100 kW.
- Memory limitations are not included in the above estimates.

## 5. References

[1] Magro A. 2013. A Real-Time, GPU-Based, Non-Imaging Back-End for Radio Telescopes. PhD Thesis, University of Malta.

[2] Clark, M. A., La Plante, P. C. and Greenhill, L. J. 2011. Accelerating Radio Astronomy Cross-Correlation with Graphics Processing Units. International Journal of High Performance Computing

[3] Dewdney, P. E. 2013. SKA1 System Baseline Design